\documentstyle[preprint,epsfig,aps]{revtex}
\draft
\preprint{ \begin{tabular}{l}
\hbox to\hsize{May, 1999 \hfill SNUTP 99-025,\ KIAS-P99029}\\
\hbox to\hsize{\hfill hep-ph/9905315}\\
\end{tabular} }
\tightenlines
\begin{document}

\def\beqar{\begin{eqnarray}}
\def\eeqar{\end{eqnarray}}
\def\beq{\begin{equation}}
\def\eeq{\end{equation}}
\def\tnu{\tilde{\nu}}
\def\tS{\tilde{S}}
\def\abs#1{\left|#1\right|}
\def\ve{\langle\tilde{\nu}_e \rangle}
\def\vmu{\langle\tilde{\nu}_{\mu} \rangle}
\def\vtau{\langle\tilde{\nu}_{\tau} \rangle}
\def\vi{\langle\tilde{\nu}_i \rangle}
\def\v2{\langle\tilde{\nu}_2 \rangle}
\def\v3{\langle\tilde{\nu}_3 \rangle}
\def\vmt{\langle\tilde{\nu}_{\mu,\tau} \rangle}
\def\vS{\langle S \rangle}
\def\vN{\langle N \rangle}
\def\vev#1{\langle#1\rangle}

\title{\Large\bf Constraints on Parameters of the $R_F$ Parity Model from 
Quark and Neutrino Mass Matrices}
\author{Jihn E. Kim$^{(a,b)}$,\footnote{E-mail: jekim@phyp.snu.ac.kr}
Bumseok Kyae$^{(a)}$, and
Jae Sik Lee$^{(b)}$\footnote{E-mail: jslee@phys.kias.re.kr} \\}
\address{$^{(a)}$Department of Physics and Center for Theoretical
Physics, Seoul National University,\\
Seoul 151-742, Korea, and\\
$^{(b)}$School of Physics, Korea Institute for Advanced Study,
Cheongryangri-dong, Dongdaemun-ku, Seoul 130-012,
Korea}
\maketitle

\begin{abstract}
The forms of quark and lepton mass matrices are severely restricted
in the $R_F$ parity model. We determine the form of the quark mass
matrix first and derive the form of the neutrino mass matrix. For 
this to be consistent with the present experiments,
we conclude that the masses of the superpartners 
of the right--handed down-type squarks and sneutrino vacuum
expectation values satisfy, ${m_{\tilde{d}_{Rj}}}\vmu \gtrsim 
400 ~{\rm GeV}^2 $ or ${m_{\tilde{d}_{Rj}}}\vtau \gtrsim 
400 ~{\rm GeV}^2 $. We also find that without a sterile neutrino
it is difficult to obtain a large mixing angle solution of
$\nu_\mu$ and $\nu_\tau$. With a sterile neutrino, we show
a possibility for a large mixing angle solution 
of $\nu_\mu$ with $\nu_{\rm sterile}$.
\end{abstract}


\newpage

\section{introduction}

The strong CP problem has two most attractive possibilities: the spontaneously
broken Peccei-Quinn (PQ) symmetry~\cite{pq} and the massless quark 
solution~\cite{choi}.
For the case of axion, one can easily construct models for a very light 
axion~\cite{axion}, since the very light axion is just an addition to
the standard model and its effect to the standard model is suppressed by 
power(s) of the axion decay constant ($F_a\sim 10^{12}$~GeV). For very light
axions, there exists a good motivation from the superstring that a
model-independent axion should exist~\cite{string}. Probably, this is
the most attractive theoretical reason for the very light axion~\cite{stax}. 
On the other hand, it has been very difficult to construct 
a consistent model for the massless up quark because any new symmetry
to make the up quark massless affects the predictions of the standard model
sector significantly. It has been conjectured that somehow a symmetry
in superstring models would render up quark massless. But this is also
disfavored by the problem of no global symmetry 
in superstring models except the one
corresponding to the model-independent axion~\cite{bd}.  
Even if one has a theory for massless up quark, it will satisfy many
constraints related to the fermion mass matrix. Here lies the difficulty
of building a massless up quark model. Therefore, it will be very
interesting to construct phenomenologically allowable massless up
quark models. Indeed, there exist an attempt to build a
model toward $m_u=0$ with a $U(1)$ horizontal symmetry~\cite{nir}.

Recently, a kind of $R$ parity named $\lq\lq R_F$ parity" was introduced 
to solve the strong CP problem with almost massless up 
quark~\cite{kkl}. The $R_F$ parity model was constructed such
that it satisfies the proton decay constraint and possibility for
generating quark and lepton masses~\cite{kkl}. In this paper, we study the
$R_F$ parity model further to constrain its parameter space from the
known quark and lepton mass matrix textures. 
We find that it is not impossible to satisfy the phenomenological
constraints.

The $R_F$ parity is defined as
\begin{equation}
R_F=(-1)^{3B+L+2S}(-1)^{2IF},
\end{equation}
where $F=\delta_{f1}\ (f=1,2,3)$, and $B, L, S, I$ and $F$ are the 
baryon number, lepton number, spin,
weak isospin, and the first family number, respectively. 
In this model the $R_F$ parity distinguishes the first
family from the second and the third ones. Thus, it is possible to make
the first family massless. This is good for up quark but bad for 
electron and down quark. Therefore, it was necessary to break the
$R_F$ parity spontaneously by vacuum expectation values (VEV) of the
sneutrino fields.
The sneutrino fields have the same quantum number
as the down type Higgs field $H_1$; thus their VEV's render electron and
down type quark massive\footnote{With a different motivation, similar idea
was considered in Ref.~\cite{liusong}.}.
Still up quark remains massless since the
quantum numbers of sneutrinos are different from the up type Higgs
field $H_2$.

Another kind of $R$ parity can be defined such that $U^c_1$ (the so-called
righthanded up quark singlet superfield) has a different quantum number
from the rest of the quark and lepton superfields. This possibility 
will make up quark massless, but does not explain why the other first
family members are light. This possibility will be discussed in Sec. V
after exploiting the phenomenological implications of the $R_F$ 
parity model.

The $R$ parities responsible for the masslessness of
up quark can have a root in superstring models. Presumably these $R$
parities can be a discrete subgroup of $U(1)_R$ global symmetry~\cite{gkn}. 
Even though the string theory does not admit a global $U(1)_R$ symmetry,
it can allow  discrete groups. Therefore, the study of discrete
groups toward massless up quark can have a far reaching extension
toward superstring models. 

In Sec. II, we discuss the textures of the quark mass matrix, 
and obtain bounds the sneutrino VEV's. In Sec. III, we study the lepton
mass matrix, and obtain bounds on the masses of zino and down type
squarks. In Sec. IV, we introduce $U(1)$ horizontal gauge symmetry
so that the desired mass matrices are obtained from the 
Froggatt-Nielsen idea~\cite{fn}.
The possibility of the opposite $R$ parity charge for $U_1^c$ is 
discussed in Sec. V. Finally, we summarize our works in Sec. VI.
In Appendix, we present superpotential terms obeying the $U(1)_X\times
U(1)_Z$ symmetry, and suggest a mechanism of generating the $\mu$-term.

\section{The CKM matrix}

The most general $d=3$ superpotential consistent with the $R_F$ parity 
is
\begin{eqnarray}
W_0=&f^l_{ij}L_iE_j^cH_1(i\ne 1)+f^u_{ij}Q_iU_j^cH_2(i\ne 1)
+f^d_{ij}Q_iD^c_jH_1(i\ne 1)\nonumber\\
&+\lambda_{1jk}L_1L_jE^c_k(j\ne 1)+\lambda^\prime_{i1k}L_iQ_1D^c_k(i\ne 1)
+\lambda^\prime_{1jk}L_1Q_jD^c_k(j\ne 1).
\end{eqnarray}
This superpotential gives $m_e=m_d=m_u=0$ naturally. 
To explain the nonzero $m_e$ and $m_d$, we introduce a
soft $R_F$ parity violating term, which is hoped to mimic a general 
feature in other $R_F$ breaking models, 
\begin{equation}
W_1=M_S S^2+\epsilon^2 S + f_{Si} SL_iH_2(i\neq1) ,
\end{equation}
where $S$ is a singlet superfield with $Y=0$ and $R_F=-1$ and
we suppress the $d=4$ baryon number violating terms.
With this $R_F$ parity violation, the electron and the down quark obtain
their masses through the vacuum expectation values (VEV's)
of the $S$ and sneutrino fields given by
\beq
\vS \sim \frac{\epsilon^2}{M_S}, ~~~~~~~
\langle \tnu_i\rangle \sim \frac{f_{Si}\,v_2\,\epsilon^2}{M^2_{\tnu_i}},
\label{nuvev}
\eeq
where $v_\alpha$'s are VEV's of the neutral Higgs fields $\langle 
H_\alpha^0\rangle\ (\alpha=1,2)$ and $M^2_{\tnu_i}$ denotes the
mass of the $i$-th generation sneutrino field.
Let us note that $\vi$'s are not suppressed by $\frac{1}{M_S}$ 
due to the $M_SS^2$ term in $W_1$.

{}From the above $R_F$ parity conserving and violating terms, the mass matrices
of up- and down-type quarks and charged leptons are given by
\beqar
(M_u)_{ij} &=& f^u_{ji} \frac{v_2}{\sqrt{2}} (j\neq 1), \nonumber \\
(M_d)_{ij} &=& f^d_{ji} \frac{v_1}{\sqrt{2}} (j\neq 1) + 
\lambda'_{1ji}\langle \tnu_1 \rangle (j\neq 1) +
\sum_{k=2,3}\lambda'_{kji}\langle\tnu_k\rangle\delta_{j1}, \nonumber \\
(M_l)_{ij} &=& f^l_{ji} \frac{v_1}{\sqrt{2}} (j\neq 1) + 
\lambda_{1ji}\langle \tnu_1 \rangle (j\neq 1) +
\sum_{k=2,3}\lambda_{kji}\langle\tnu_k\rangle\delta_{j1}. 
\label{massmtx}
\eeqar
The vanishing elements of the
first column of mass matrix of the up-type quark guarantees the massless up
quark at the tree level. And
it is obvious that the elements of the
first columns of mass matrices of the down-type quarks
and the charged leptons are also zero without the soft $R_F$ parity 
violation, i.e. without sneutrino VEV's,
$\langle\tilde{\nu}_2 \rangle$ and $\v3$.

Here, we try to satisfy the phenomenologically known quark
mass hierarchy. This hierarchy is expressed in terms of a small
expansion parameter, $\lambda\simeq 0.22$ which is sine of the
Cabibbo angle. Using renormalization group equations,\footnote{Only 
the strong interaction coupling is dominant at energy scales 
lower than the top quark mass.} we have
\beq
\frac{m_d(m_t)}{m_b(m_t)}\sim\lambda^4, ~~~
\frac{m_s(m_t)}{m_b(m_t)}\sim\lambda^2, ~~~
\frac{m_c(m_t)}{m_t(m_t)}\sim\lambda^4, ~~~
\frac{m_b(m_t)}{m_t(m_t)}\sim\lambda^3.
\eeq
These ratios of masses indicate that the unitary matrix defining
mass eigenstate up type quarks in terms of the weak eigenstate quarks 
has the form $U_L={\bf 1}+{\cal O}(\lambda^4)$ up to
phases.\footnote{We find that the phases and ${\cal O}(\lambda^4)$ term
of $U_L$ do not change our results.} Therefore, we consider only the 
down-type quark mixing from now on, or $V_{\rm CKM} = D_L +{\cal O}
(\lambda^4)$.

In case that only down-type quarks mix, $M_d$ and 
$V_{\rm CKM}$ are constrained by the following relation,
\beq
V_{\rm CKM}^{\dagger}M_d^{\dagger}M_d V_{\rm CKM}
={\rm diag}(m_d^2,m_s^2,m_b^2) 
\label{mdckm}
\eeq
which allows us to extract constraints on the parameters of $M_d$
from the measured values of the $V_{\rm CKM}$ elements.  

Another experimental data we take into account in the $R_F$ model are the
decay modes of $K^+$~\cite{kpibound},
\begin{equation}
{\cal B}(K^+ \rightarrow \pi^+ \nu \bar{\nu})\le 5.2\times 10^{-9},
\ \ \ {\cal B}(K^+ \rightarrow \pi^0 \nu e^+)\le 0.0482,
\end{equation} 
from which one can derive a bound on $\lambda'_{(i\neq 1)1j}$~\cite{kpibound},
\beq
\abs{\lambda'_{(i\neq 1)1j}} \lesssim 0.012 \,
\left( \frac{ m_{\tilde{d}_{Rj}} }{100~{\rm GeV}} \right),
\label{kpi}
\eeq
where $m_{\tilde{d}_{Rj}}$ denotes the right--handed down--type squark mass.
Combining this upper bound with constraints from $V_{\rm CKM}$ will
lead us to lower bounds on the sneutrino VEV's.

If $R_F$ parity is exact, then
the mixing matrices of up- and down-type left-handed quarks, which
diagonalize $M_u^{\dagger}M_u$ and$M_d^{\dagger}M_d$, are given by~\footnote{
The matrices $U_L,D_L,u_L$ and $d_L$ should not be confused with the particle
names with the same notations.}
\beq
U_L=\left[\begin{array}{cc}
1 & 0  \\
0 & u_L
\end{array}\right] ,
~~~~~~~~
D_L=\left[\begin{array}{cc}
1 & 0  \\
0 & d_L
\end{array}\right], 
\label{mixworf}
\eeq
where $u_L$ and $d_L$ are $2\times 2$ matrices. {}From these, 
the CKM matrix is given by
\beq
V_{\rm CKM}^{^{R_F}}=
\left[\begin{array}{cc}
1 & 0  \\
0 & u_L^{\dagger}d_L
\end{array}\right] .
\eeq
Therefore, mixing of the first family with the other families
and a CP phase do not appear in $V_{\rm CKM}^{R_F}$. Thus the $R_F$ parity
must be broken spontaneously to render phenomenologically
acceptable angles. The spontaneous symmetry breaking of the
$R_F$ parity is achieved by the VEV's of sneutrino fields,
$\vi~ (i=2,3)$.

In the presence of sneutrino VEV's, let
us parameterize the down-type quark mass matrix as
\beq
M_d=
\left[\begin{array}{ccc}
\epsilon_1 & (m_2)_1 & (m_3)_1 \\
\epsilon_ 2& (m_2)_2 & (m_3)_2 \\
\epsilon_3 & (m_2)_3 & (m_3)_3 
\end{array}\right] \equiv (\vec{\epsilon}, \vec m_2,
\vec m_3).
\eeq
Then $M^{\dagger}_dM_d$ is given by
\beq
M^{\dagger}_dM_d=\left[\begin{array}{ccc}
|\vec{\epsilon}|^2 & \vec \epsilon^*\cdot \vec m_2 &
\vec \epsilon^*\cdot \vec m_3  \\
    \vec{\epsilon}\cdot \vec m_{2}^*  & |\vec m_2|^2 &
\vec m_{2}^*\cdot \vec m_3\\
    \vec{\epsilon}\cdot \vec m_{3}^*  & \vec m_{2}\cdot \vec m_{3}^* &
|\vec m_3|^2
\end{array}\right].
\eeq
{}From Eq.~(\ref{mdckm}), we obtain the following relations,
\begin{eqnarray}
|\vec{\epsilon}|&=&\left(\sqrt{1+A^2|z|^2}+{\cal O}(\lambda)
\right)\frac{m_d}{\lambda}
\approx \sqrt{1+A^2|z|^2}\lambda m_s \, , \nonumber \\
|\overrightarrow{m_2}|&=&\left(\sqrt{1+A^2}+{\cal O}(\lambda)\right)m_s \, ,
\nonumber \\
|\overrightarrow{m_3}|&=&\left(1+{\cal O}(\lambda^3)\right)m_b \, , \nonumber \\
\vec{\epsilon^*}\cdot \overrightarrow{m_2}&=&
|\vec{\epsilon}||\overrightarrow{m_2}|
\left(\frac{1+zA^2}{\sqrt{(1+A^2)(1+A^2\abs{z}^2)}}+{\cal O}(\lambda)\right)
\, , \nonumber \\
\vec{\epsilon^*}\cdot \overrightarrow{m_3}&=&
|\vec{\epsilon}||\overrightarrow{m_2}|
\left(\frac{zA}{\sqrt{1+A^2\abs{z}^2}}+{\cal O}(\lambda)\right) \, , \nonumber \\
\overrightarrow{m_{2}}\cdot \overrightarrow{m_3}&=&
|\overrightarrow{m_2}| |\overrightarrow{m_3}|
\left(\frac{A}{\sqrt{1+A^2}}+{\cal O}(\lambda)\right) \, , 
\end{eqnarray}
where $z=\rho-i\eta$ and $\lambda$, $A$, $\rho$, and $\eta$ are the 
conventional Wolfenstein parameters~\cite{Wolf}. Therefore,
for example, one can obtain the following down-type quark mass matrix which
has eigenvalues of $m_d$, $m_s$ and $m_b$ and gives a right form for
the CKM mixing matrix
\beq
M_d=\left[\begin{array}{ccc}
{\cal O}(\lambda^4)& {\cal O}(\lambda^3) & {\cal O}(\lambda^4) \\
     \lambda^3 &\lambda^2 &{\cal O}(\lambda^3) \\
     z^*A\lambda^3 &A\lambda^2 & 1
\end{array}\right] m_b \, .
\label{md}
\eeq
{}From Eqs. (\ref{massmtx}) and (\ref{md}), one can derive the following
relations 
\beqar
\lambda'_{211}\vmu +\lambda'_{311}\vtau &=& {\cal O}(\lambda^4)~m_b,
\nonumber \\
\lambda'_{212}\vmu +\lambda'_{312}\vtau &=& \lambda^3~m_b,
\nonumber \\
\lambda'_{213}\vmu +\lambda'_{313}\vtau &=& z^*A\lambda^3~m_b.
\label{res1}
\eeqar
If one of $\vmu$ or $\vtau$ dominates, the relative sizes of 
$\lambda'$'s can be fixed independently of the sizes of
the sneutrino VEV's.  The numerical values of the CKM matrix elements 
are given for $\lambda =0.22$ and $\abs{z^* A}=0.34$ \cite{ckm}. 
Therefore, a typical size of the
products $\lambda'_{i1j}\vi$ with $i=2,3$ and $j=1,2,3$ 
is $(0.01-0.05)$ GeV.
On the other hand, it is not easy to derive some meaningful 
information on the other $\lambda'_{1(i\neq 1)j}$'s 
because these couplings contribute to $M_d$ together with 
the conventional Yukawa terms proportional to
unknown parameters $f^d_{ij}$ and $v_1$.
{}From Eqs.~(\ref{kpi}) and (\ref{res1}), it is obvious that at least
one of $\vmu$ and $\vtau$ should satisfy the following
inequality :
\beq
\vmt \gtrsim 4 ~{\rm GeV}
\left( \frac{100~{\rm GeV}}{m_{\tilde{d}_{Rj}}} \right).
\label{lower}
\eeq
Note that the lower bound becomes smaller for a larger squark mass.

Though the up--quark mass is zero at tree level, it can be generated
radiatively when $S$ field has a vacuum expectation value as shown in Fig. 1
\cite{kkl}.
The one-loop up--quark mass is given by
\begin{equation}
\delta m_u \sim \sum_ {i,k=2,3}
\frac{f_{k1}^{u}\lambda'_{i1k}}{16\pi^2}
\frac{m^{(d)}_k f_{Si} \langle S\rangle}{M_{\rm SUSY}}
\sim \sum_{i=2,3}
\frac{f_{Si}\lambda'_{i13} \epsilon^2}{16\pi^2}
\frac{m_b f_{31}^{u}}{M_{\rm SUSY} M_S} \, ,
\end{equation}
where $m^{(d)}_2=m_s$ and $m^{(d)}_3=m_b$ and we used $\langle S\rangle =
\epsilon^2/M_S$.
The combination
$f_{Si}\lambda'_{i13} \epsilon^2$ is constrained as in Eq.~(\ref{res1})
through the relation 
$\langle \tnu_i\rangle \sim {f_{Si}\,v_2\,\epsilon^2}/{M^2_{\tnu_i}}$.
Taking $m_b=5$ GeV, $v_2 = 100$ GeV  and $M^2_{\tnu_i} = 1$ TeV gives 
$f_{Si}\lambda'_{i13} \epsilon^2 \sim 180$ GeV$^2$. This one--loop
up--quark mass should be small to solve the strong CP problem :
$\delta m_u < 10^{-13}$ GeV \cite{stax}. This leads to
\beq
\frac{f_{31}^{u}}{(M_S/{\rm GeV})}< 2 \times 10^{-11} .
\eeq
This bound is stronger than the rough estimation
given in Ref.~\cite{kkl} by one order of magnitude.

\section{Neutrino mixing and sterile neutrino}

Due to the VEV's of the sneutrino and $S$ fields in our case, 
three neutrinos mix with four
neutralinos and one $S$ field in the $8\times 8$ mass matrix.
In the $(\nu_i;\tilde{B},\tilde{W}_3,\tilde{H}_1^0,\tilde{H}_2^0,S)$ basis,
the mass matrix is given by
\beq
M_0=
\left(
\begin{array}{cc}
0 & m_D \\
m_D^T & M
\label{m88}
\end{array}
\right),
\eeq
where we neglected the contributions coming from the one-loop
diagrams.
Here, 0 is the $3\times 3$ matrix with 0 entries and  $m_D$ is a 
$3\times 5$ matrix
\begin{equation} \label{m26}
m_D=\left(\begin{array}{ccccc}
-{1 \over 2}g' \ve   & {1 \over 2}g \ve    & 0 & -\mu_1 & 0 \\
-{1 \over 2}g' \vmu  & {1 \over 2}g \vmu  & 0 & -\mu_2 & 
f_{S2} \frac{v_2}{\sqrt{2}}\\
-{1 \over 2}g' \vtau & {1 \over 2}g \vtau & 0 & -\mu_3 &  
f_{S3} \frac{v_2}{\sqrt{2}}
\end{array}\right) ,
\end{equation}
where $\mu_{2,3}=f_{S(2,3)}\vS$. $M$ in Eq.~(\ref{m88}) is a 
$5 \times 5$ mass matrix of the neutralinos and $S$ field
\begin{equation} \label{m27}
M=\left(\begin{array}{ccccc}
cM_2 & 0 & -{1 \over 2}g'v_1 & {1 \over 2}g'v_2 & 0 \\
0 & M_2 & {1 \over 2}gv_1 & -{1 \over 2}gv_2 & 0 \\
-{1 \over 2}g'v_1 & {1 \over 2}gv_1 & 0 & - \mu & 0 \\
{1 \over 2}g'v_2 & -{1 \over 2}gv_2 & - \mu & 0 & (f_{S2}+f_{S3})\vS \\
0 & 0 & 0 & (f_{S2}+f_{S3})\vS & M_S
\end{array}\right) ,
\end{equation}
where $c\equiv M_1/M_2=(5/3)\tan^2\theta_W\simeq 0.5$, 
assuming the unification relation. 

Since a typical scale for $M$ is much larger than 
that of $m_D$, it is enough to use the
see-saw formula to find the following reduced neutrino mass matrix \cite{pil}
:
\begin{eqnarray} 
m_{\rm eff}^{\nu}&=& -m_D\; M^{-1}\; m_D^T  \nonumber \\
&=& {cg^2+ g'^2 \over D}\,
\left(\begin{array}{ccc}
\Lambda_e^2 & \Lambda_e \Lambda_\mu
& \Lambda_e \Lambda_\tau \\
\Lambda_e \Lambda_\mu & \Lambda_\mu^2
& \Lambda_\mu \Lambda_\tau \\
\Lambda_e \Lambda_\tau & \Lambda_\mu \Lambda_\tau & \Lambda_\tau^2
\end{array}\right) 
-\frac{v_2^2}{2M_S}
\left(\begin{array}{ccc}
0 & 0 & 0 \\
0 & f_{S2}^2 & f_{S2}f_{S3}  \\
0 & f_{S2}f_{S3}& f_{S3}^2 
\end{array}\right) 
+{\cal O}\left(\frac{1}{M_S^2}\right),
\label{mnu}
\end{eqnarray}
where $\Lambda_i$ and $D$ are given by
\beqar
\Lambda_i&=&\mu\vi-v_1\mu_i\, , \nonumber \\
D &=& 2\mu \left[-2cM_2\mu + v_1v_2
\left(cg^2 + g'^2 \right)\right].
\eeqar

Neglecting the ${\cal O}(1/M_S^2)$ contributions,
the neutrino mass matrix, Eq.~(\ref{mnu}), 
has one zero eigenvalue and two nonzero eigenvalues 
\beqar
\label{numass}
m_{\nu_1}&=& 0,\nonumber\\
m_{\nu_2}&=&
-\frac{\left[(f_{S2}\Lambda_{\tau}-f_{S3}\Lambda_{\mu})^2+
(f_{S2}^2+f_{S3}^2)\Lambda_e^2\right]v_2^2}{2\Lambda^2 M_S} ,
\\
m_{\nu_3}&=&
\frac{(cg^2+g'^2) \Lambda^2}{D}
-\frac{(f_{S2}\Lambda_{\mu}+f_{S3}\Lambda_{\tau})^2v_2^2}
{2\Lambda^2 M_S},
\nonumber
\eeqar
where $\Lambda^2=\Lambda_e^2+\Lambda_{\mu}^2+\Lambda_{\tau}^2$.  

Naively, one would expect that
$\Lambda_{\mu,\tau}$ are smaller than $\Lambda_e$ because
$\Lambda_e$ comes from the $R_F$-parity conserving parts. But, this
expectation is not true
because $\vmt$ given in Eq.~(\ref{nuvev}) is not suppressed by $M_S$.
Namely, the sneutrino $\tilde \nu_e$,
preserving the $R_F$ parity, 
and sneutrinos $\tilde\nu_{\mu,\tau}$, violating the $R_F$ parity, 
can have comparable VEV's, or even $\ve$ can be smaller than
$\vmt$ depending on the sizes of the corresponding soft SUSY breaking terms. 
This is because below the $R_F$ breaking scale
their VEV's are determined by the respective potentials. 
Then, the mixing matrix $V_\nu$ which diagonalizes
$m_{\rm eff}^{\nu}$ satisfies 
$V_{\nu}^ T\, m_{\rm eff}^{\nu}\, V_{\nu} =
{\rm diag}(m_{\nu_1},m_{\nu_2},m_{\nu_3})$, and
has the following form
\beq
V_{\nu} =
\left(
\begin{array}{ccc}
\Lambda_{\tau}/\Lambda'
& -(\Lambda_e\Lambda_{\mu})/(\Lambda\Lambda') & \Lambda_e/\Lambda \\
0 & \Lambda'/\Lambda  & \Lambda_{\mu}/\Lambda  \\
-\Lambda_e/\Lambda' 
& -(\Lambda_{\mu}\Lambda_{\tau})/(\Lambda\Lambda')& \Lambda_{\tau}/\Lambda  \\
\end{array}
\right)+{\cal O}(\frac{1}{M_S}),
\eeq
where 
$\Lambda'\,^2\equiv\Lambda_{e}^2+\Lambda_{\tau}^2$.
On the other hand, the mixing matrix of the left--handed
charged leptons\footnote{The matrix $L_L$ should not be confused with
the left-handed lepton superfield.} ($L_{L}$)
which satisfies 
$L_{L}^{\dagger}M_l^{\dagger}M_lL_{L}={\rm diag}(m_e,m_{\mu},m_{\tau})$
does not mix the first family with the other family
members without $R_F$--parity
violation as the left--handed quark mixing matrices in Eq.~(\ref{mixworf})
do not mix the first family members.
Therefore, in the presence of $R_F$ parity violation we obtain 
\beq
\nu_e = \left(L_L^{\dagger}V_{\nu}\right)_{1k}\nu_{k}
\approx 
(\Lambda_{\tau}/\Lambda')~\nu_1-
(\Lambda_e\Lambda_{\mu})/(\Lambda\Lambda')~\nu_2+
(\Lambda_e/\Lambda)~\nu_3 
\eeq
where $\nu_i\ (i=1,2,3)$ are the mass eigenstates.

{}From the lower bounds on $\vmt$ given in Eq.~(\ref{lower}), one can 
obtain the following lower bound on $m_{\nu_3}$ :
\beq
m_{\nu_3} \ge  \frac{cg^2+g'^2}{D}\mu^2\vmt^2 \sim
~80~{\rm KeV}~
\left(\frac{300~{\rm GeV}}{M_2} \right)~
\left(\frac{1~{\rm TeV}}{m_{\tilde{d}_{Rj}}} \right)^2
,
\eeq
where we 
take $c=0.5$ as given by the unification condition.
This means that $\Lambda_{\mu}$ and/or $\Lambda_{\tau}$ are 
significantly greater than $\Lambda_e$ to avoid the present mass bound on
the  electron neutrino: $m_{\nu_e} < {\cal O}(1)$ eV.

Let's assume  
$\Lambda_{\tau} \gg \Lambda_e$ and the off-diagonal elements of
charged lepton mass matrix $M_l$ 
are zero, then we obtain
\beqar
\nu_e &\sim &  \nu_1 \, , \nonumber \\
\nu_{\mu} &\sim & \left(
\Lambda_{\tau} \nu_2 + 
\Lambda_{\mu}\nu_3\right)/\Lambda \, , \nonumber \\
\nu_{\tau} &\sim & \left(
-\Lambda_{\mu}\nu_2 +
\Lambda_{\tau}\nu_3\right) /\Lambda \, .
\eeqar
If one of $\Lambda_\mu$ and $\Lambda_\tau$ dominates, then
$\nu_\mu$ or $\nu_\tau$ will be the mass eigenstate $\nu_3$
whose mass is greater than $\sim 100$ KeV.
This implies that it is hard to explain
the observation of the large mixing and
$\sqrt{\Delta_{\rm atm}^2}\simeq 5\times 10^{-2}$ eV
given by the Super-Kamiokande in terms of $\nu_\mu - \nu_\tau$ oscillation 
\cite{kami}. To explain the Super-Kamiokande observation, let's
introduce the sterile neutrino $N$ which is a singlet superfield with $Y=0$
and $R_F=-1$.  The relevant additional superpotential is given by
\beq
W_2=M_N N^2+M_{NS} N S + f_{N_i}NL_iH_2(i\neq 1)
+\frac{\tilde\lambda''_{Nijk}}{M_{_P}}N U^c_i D^c_j D^c_k,
\eeq
where $M_{_P}$ is the Planck mass. 
The VEV of $N$ field $\vN$ induced by $\vS$ is
\beq
\vN \sim \frac{\epsilon^2 M_{NS}}{M_{\tilde N}^2}
+\frac{\epsilon^2 M_N M_{NS}}{M_S M_{\tilde N}^2}
+\frac{(f_{S_i}+f_{N_i})M_N \vi v_2}{M_{\tilde N}^2} .
\eeq
If $M_N$ and $M_{NS}$ are negligible, then $\vN\simeq 0$
and hence we can avoid the fast proton decay,
arising from the last term of Eq.~(28).
The explicit mixing between $S$ and $N$ gives two mass eigenstates $S'$ and
$N'$ with
\beq
M_{S'}\approx M_S~~{\rm and}~~M_{N'}\approx M_N-M_{NS}^2/M_S.
\eeq
With appropriate $M_N$ and $M_{NS}$ values, one could obtain
almost vanishing sterile neutrino mass $M_{N^\prime}$.

Assuming $|m_{\nu_{\mu}}^2-m_{\nu_{_N}}^2| \simeq \Delta_{\rm atm}^2$ 
and $\Lambda_{\tau} \gg \Lambda_{\mu}$, 
from Eqs.~(\ref{nuvev}) and (\ref{numass}), we obtain
\beq
m_{\nu_{\mu}} \sim m_{\nu_2} \sim \frac{f_{S2}^2 v_2^2}{2M_S}
\sim \frac{M_{\tnu_2}^4 \vmu^2}{2\epsilon^4 M_S}.
\eeq
Taking $\sqrt{\Delta_{\rm atm}^2}\simeq m_{\nu_{\mu}}$ assuming almost
vanishing sterile neutrino mass, we obtain
\beq
M_S \sim 10^{10}~{\rm GeV}~~
\left(\frac{M_{\tnu_2}}{\epsilon}\right)^4
\left(\frac{\vmu^2}{1~{\rm GeV}^2}\right) .
\eeq
For $M_{\tilde \nu_2}\simeq 10$~TeV, $\epsilon\simeq 100$~GeV, and
$\vmu\simeq 10^{-3}$~GeV, the mass parameter $M_S$ should be around
$10^{12}$~GeV which is the intermediate scale needed in
supergravity and invisible axion~\cite{int}.
Also, the large mixing between $N$ and
$\nu_{\mu}(\sim\nu_2)$ could be obtained taking 
$f_{N_2} v_2 \approx m_{\nu_2}$  with $f_{N_2}\gg f_{N_3}$.

Note that the assumed hierarchy between the sneutrino VEV's
$\vtau \gg \vmu , \ve $ is consistent with the neutrino data
since we introduced a sterile neutrino. In this case, $\nu_\tau$ can
decay to a lighter neutrino plus photon, which may affect the
evolution of the universe. 
The neutrino interaction
with electromagnetic fields due to nonzero transition magnetic moment is
described by 
\beq
f'_i\mu_B(\bar\nu_i\sigma_{\mu\nu}\nu_\tau) F^{\mu\nu}+{\rm H. c.}\, ,
\eeq
where $\mu_B$ denotes the Bohr magneton.
Then the partial lifetime of the tau neutrino is
\cite{gni}
\begin{equation}
\tau_{\nu_\tau\rightarrow\nu_i}\simeq
\frac{2\pi}{m_{\nu_\tau}^3 f^{\prime2}_i \mu_B^2}
\approx  4.5 \times 10^{-3}
\left(\frac{100~{\rm KeV}}{m_{\nu_\tau}}\right)^3
\left(\frac{10^{-7}}{f^\prime_i}\right)^2
~{\rm sec}\, .
\end{equation}
The decay, $\nu_\tau\rightarrow
\nu_i+\nu_j+\bar\nu_k$, is negligible compared to the above photon
mode.
 
Note also that our assumed hierarchy on the sneutrino masses that
the third generation VEV is much smaller than those
of the first and second ones, $M_{\tnu_3} \ll M_{\tnu_{1,2}}$, is possible
in the so-called effective SUSY framework~\cite{effsusy}.\footnote{This 
effective SUSY was proposed to suppress the SUSY
contributions to low-energy flavor changing neutral current (FCNC) or
CP violating processes.}

\section{Horizontal symmetry}

The mass hierarchy of quarks and charged
leptons can be explained by  introducing an {\it abelian horizontal 
symmetry }$U(1)_X \times U(1)_Z$ in our framework. This kind
of horizontal symmetry at high energy scale, presumably at
the Planck scale, might be necessary to introduce an expansion
parameter in the mass matrix~\cite{fn}. In string models,
there appear many gauge $U(1)$ symmetries which act as
horizontal symmetries in models distinguishing all
fixed points~\cite{ckn}. In this case,
the expansion parameter(s) is the ratio(s)
of VEV(s) of SM singlet field(s) and the string scale. 
In this spirit, we try to search for a possibility of introducing
the mass matrices discussed in the previous sections.

Because it is very difficult to study the general mass matrix,
we are guided first by the phenomenologically plausible
relation ${M_l}\approx {M_d}$ (but with a modification {\ a la}
Georgi and Jarlskog~\cite{gj}), and a phenomenological hierarchy 
${m_c^2(m_t)}/{m_t^2(m_t)}\sim\lambda^8$. As explained before,
then $\abs{U_L}={\bf 1}+{\cal O}(\lambda^4)$; and 
we can take the following forms for the mass matrices 
using $\lambda$ as an expansion parameter

\beq
{M_u}\sim \left[\begin{array}{ccc}
0 &\lambda^6 &\lambda^2\\
     0 &\lambda^5 &\lambda  \\ 
0 &\lambda^4  &  1      
\end{array}\right] m_t \, ,~~~
{M_d}\sim \left[\begin{array}{ccc}
\lambda^7& \lambda^6 & \lambda^7 \\ 
     \lambda^6 &\lambda^5 &\lambda^6 \\   
     \lambda^6 &\lambda^5 &\lambda^3   
\end{array}\right] m_t \, ,~~~
{M_l}\sim \left[\begin{array}{ccc}
 \lambda^7 &\lambda^6 &\lambda^7\\ 
     \lambda^6 &\lambda^5 &\lambda^6\\ 
     \lambda^6 &\lambda^5 &\lambda^3 
\end{array}\right] m_t \, .
\label{mm}
\eeq
Some elements arise as higher order corrections. For example,
the first and the second rows of $M_u$ and (1,i) and (2,3) 
elements of $M_d$ arise as higher order corrections. Note that
the lower right $2\times 2$ submatrix of $M_u$ gives a
determinant of order $\lambda^5m_t^2$. However, this does not
imply $m_c\sim \lambda^5m_t$ since $M_u$ must be diagonalized by
a biunitary transformation, implying this determinant is not
invariant. To get eigenvalues, we diagonalize $M_uM_u^\dagger$
in which case we obtain $m_c\sim \lambda^4m_t$.

Let us introduce three SM singlet fields,
$\theta_{1}$, $\theta_{2}$ and $\theta_{3}$, to explain the above
mass hierarchies, which seems to be our minimal choice.
We further assume a universal VEV to simplify
the analysis,
\beq
\langle\theta_1\rangle ^3\approx
\langle\theta_2\rangle ^3\approx
\langle\theta_3\rangle ^3\approx
\lambda M^3, 
\eeq
where $M$ is the energy scale where nonrenormalizable
interactions are introduced but still
the $U(1)_X \times U(1)_Z$ symmetry is preserved
below $M$. In string models, one can identify $M=M_{\rm string}$.

We find that the following $U(1)_{X,Z}$ charge assignments are enough to
explain the hierarchies of the mass matrices given in Eq. (\ref{mm}) 

\begin{center}
\centerline{Table 1. $U(1)_X\times U(1)_Z$ charges.}
\vskip 0.2cm
\begin{tabular}{|c|ccc|ccc|ccc|}
\hline
\vphantom{$\frac{A}{B}$} 
& $~~~Q_1~~~$ & $~~~Q_2~~~$ & $~~~Q_3~~~$ & $~~D_1^c~~$ & $~~D_2^c~~$ &
$~~D_3^c~~$ & 
   $~U_1^c~$ & $~U_2^c~$ & $~U_3^c~$ 
\\ \hline 
$Q_{_X}$ & $9/2$ & $0$ & $-5$ & $-2$ & $-2$ & $4$ & 
        $1/3$ & $1/3$ & $1/3$ 
\\ \hline
$Q_{_Z}$ & $6$ & $-1$ & $1$ & $-9$ & $-6$ & $-3$ & 
        $-1$ & $2$ & $5$ 
\\ \hline
\end{tabular}

\bigskip

\begin{tabular}{|c|ccc|ccc|ccc|}
\hline
 & $~~~L_1~~$ & $~~~L_2~~$ & $~~~L_3~~$ & 
   $~E_1^c~$ & $~E_2^c~$ & $~E_3^c~$ & $~~H_1~~$ & $~~H_2~~$ & $~~~S~~~$ 
\\ \hline
$Q_{_X}$ & $      4$ & $-1/2   $ & $  -11/2$ & 
        $-3/2$ & $-3/2$ & $    9/2$ & 
        $     -1$ & $14/3$ & $      x$ 
\\ \hline
$Q_{_Z}$ & $2$ & $-5$ & $-3$ & 
        $-5$ & $-2$ & $1$ & $-5$ & $-6$ & $z$ 
\\ \hline
\end{tabular}
\end{center}
where $Q_{_X}$ and $Q_{_Z}$ denote the $U(1)_X$ and $U(1)_Z$ charges,
respectively.  
The above charge assignment is valid for $\tan\beta \sim {\cal O}(1)$.
For large $\tan\beta$, a somewhat different charge assignment will be 
obtained. Here, we have not fixed the charges of the $S$ field.
Its charges can be fixed once we know the size of $M_S/M$.
One has to cancel anomalies. Since we are considering a
low energy effective theory, we will ignore the $U(1)_{X,Z}$
related anomalies except $U(1)_{X,Z}-SU(3)_c-SU(3)_c$ anomaly.
This is because an introduction of colored scalars at intermediate scale
is dangerous for proton stability. If the symmetry breaking scale of
$U(1)_{X,Z}$ is near the grand unification scale, then
the proton stability problem reduces to that of supersymmetric 
models~\cite{wy}. Anyway, we assume that there is no colored scalars
below the scale $M$. The other anomalies can be canceled by
introducing $Q_X$ and $Q_Z$ carrying color singlet superfields which
we do not try to specify due to our ignorance of the super 
high energy physics.  Note that as in any hierarchical 
model of this type, the ratio of charges of horizontal
symmetry $U(1)_{X,Z}$ can include a large number.

The Yukawa couplings rendering quark masses are of the
form $Q_i(D^c\ {\rm or\ }U^c)H_\alpha$ ($\alpha$ = 1 or 2).
Among these, only $Q_3U_3^cH_2$ is allowed by 
the $SU(3)_c\times SU(2)_L\times U(1)_Y\times U(1)_X\times U(1)_Z$
symmetry. To introduce other Yukawa couplings through nonrenormalizable
terms, we need at least three singlet superfields $\theta_1$, 
$\theta_2$ and $\theta_3$ which are assigned with the charges:
\begin{center}
\centerline{Table 2. $U(1)_X\times U(1)_Z$ charges of superheavy singlets.}
\vskip 0.2cm
\begin{tabular}{|c|ccc|}\hline
  &$\theta_1$ &$\theta_2$ &$~\theta_3$ \\ \hline 
$Q_{_X}$ &$0$ &$-1$ &$~1$ \\
$Q_{_Z}$ &$1$ &$-1$ &$~0$ \\ \hline 
\end{tabular}
\end{center}
Then the needed SM invariants become neutral through (non)renormalizable 
interaction with the $\theta_i$'s.  
The above charge assignments do not allow the
interactions which can give the nonzero elements 
in the first column of $M_u$.
All the other elements of $M_u$ and $M_{d,l}$ anticipated in
Eq. (\ref{mm}) can be obtained through 
interactions which are
neutral under the  $U(1)_X \times U(1)_Z$ horizontal symmetry.
These interactions are shown in the Appendix.

Note that the above charge assignment suggests the relation
$\vev{\theta_1}=\vev{\theta_2}=\vev{\theta_3}$ since only the combination of
$\theta_1\theta_2\theta_3$ is neutral. 
The superpotential of $\theta$ fields is given by
\beq
W_{\theta}=\lambda_1(\theta_1\theta_2\theta_3)
+\lambda_2(\theta_1\theta_2\theta_3)^2+\cdot \cdot \cdot .
\eeq
Since this superpotential is symmetric under the exchanges of (1,2,3),
it may hints the same VEV's of $\theta_1$, $\theta_2$,
and $\theta_3$. But at this stage only the hyperplane $\theta_1
\theta_2\theta_3$ is guaranteed. 
The D-flat conditions are
\beq
\sum_{a=1}^{3} Q^a_{_X}\vert\theta_a\vert^2=0 , ~~~
\sum_{a=1}^{3} Q^a_{_Z}\vert\theta_a\vert^2=0 ,
\eeq
where $Q^a_{_X}$ 
and $Q^a_{_Z}$ are the $U(1)_{X}$ and
$U(1)_{Z}$ charges of $\theta_a$, respectively.  Therefore, 
${\theta_1}={\theta_2}={\theta_3}$ is the D-flat direction.
In addition, the introduction of a
universal soft term in the potential $m_{3/2}^2(\theta_1^2+
\theta_2^2+\theta_3^2)$ would not destroy
the relation $\langle \theta_1\rangle=
\langle\theta_2\rangle=\langle\theta_3\rangle$.
{}From this consideration, we conclude that $\theta$ 
fields have the same magnitude of the vacuum expectation values 
\begin{equation}
\langle\theta_1\rangle=\langle\theta_2\rangle=\langle\theta_3\rangle \equiv
\lambda^{\frac{1}{3}}M \  
\end{equation}
where $\lambda$ is the expansion parameter. Let us consider
the (2,2) element of $M_u$ of Eq.~(\ref{mm}) as an illustration. 
The relevant term which is neutral under the 
horizontal symmetry is
\beq
\Theta_1^{10} \Theta_2^5  Q_2  U_2^c  H_2,
\eeq
where the dimensionless parameter $\Theta_i$ are defined as
$\Theta_i\equiv\theta_i/M$ and we assume an ${\cal O}(1)$
coupling. The VEV of $\theta_i$ of order of $\lambda^{1/3}$ and   
$\vev{H_2}=v\sin\beta$ give
\beq
\lambda^5 v \sin\beta U_2 U_2^c \approx \lambda^5 m_t U_2 U^c_2
\eeq
which is the ${\cal O}(\lambda^5)$ given in Eq.~(\ref{mm}).
Similarly, other elements of $M_u$
can be obtained through interactions which are
neutral under the  $U(1)_X \times U(1)_Z$ horizontal symmetry.

Now, let us consider the down type quark mass matrix $M_d$. 
For example, the terms which give the (1,1) element
of $M_d$ are
\beq
\Theta_1^6   \Theta_3^{3}L_3Q_1D_1^c~~{\rm and}~~
\Theta_1^{10}\Theta_2^{2}   L_2Q_1D_1^c
\eeq
which are responsible for a nonvanishing down quark mass.
After $\Theta_i$'s develop VEV's of order of $\lambda^{1/3}$ and 
assuming 
$ \langle\tilde{\nu}_2 \rangle\approx 0$ and $\v3 \approx \lambda^4 m_t$,
we obtain
\beq
\lambda^3 \v3 D_1 D_1^c \approx \lambda^7 m_t  D_1 D_1^c.
\eeq
Similarly, other elements of $M_u$
could be obtained through interactions which are
neutral under the  $U(1)_X \times U(1)_Z$ horizontal symmetry.
Of course, the above dimensionless coupling constants are given 
at very high energy scale, e.g. at $M$.

Note that our $U(1)_X\times U(1)_Z$ symmetry 
includes the result of the $R_F$ parity model~\cite{kkl}, namely 
the $U(1)_X$ among $U(1)_X\times U(1)_Z$ does not allow $R_F$ 
parity violating terms. But the $U(1)_X\times U(1)_Z$ is more
restrictive than the $R_F$ parity: it dictates the fermion 
mass hierarchy through the mass matrix Eq.~(\ref{mm}).

\section{The Opposite $R$ Parity Charge for $U^\protect{c}_1$}

Another choice of making the up quark massless is to distinguish
$U_1^c$ from the others. Thus, an
$R$ parity called $R_U$ can be defined such that only $U^c_1$ , $H_1$ and
$H_2$ have different quantum numbers
from the rest of the quark and lepton superfields.
For this purpose, let us assign the
$R_U$ quantum numbers as 
\beq
R_U(U^c_1)=R_U(H_1)=R_U(H_2)=+1~~{\rm and}~~ R_U({\rm other~fields})=-1.
\eeq
Then, the $R_U$-parity conserving superpotential is given by
\begin{eqnarray}
W_{R_U}=f^l_{ij}L_iE_j^cH_1+f^u_{ij}Q_iU_j^cH_2(j\ne 1)
+f^d_{ij}Q_iD^c_jH_1+\lambda''_{ij} U_1^c D_j^c D_k^c + \mu H_1 H_2,
\end{eqnarray}
where $\lambda''_{ij}=-\lambda''_{ij}$ . This superpotential gives the
following $Q_{em}=2/3$ and $Q_{em}=-1/3$ quark mass
matrices,
\beq
M^{(2/3)}=\left(
\begin{array}{ccc}
0&0&0 \\ H_2^0&H_2^0&H_2^0 \\ H_2^0&H_2^0&H_2^0 
\end{array}
\right),~~~~~~
M^{(-1/3)}=\left(
\begin{array}{ccc}
H_1^0&H_1^0&H_1^0 \\ H_1^0&H_1^0&H_1^0 \\ H_1^0&H_1^0&H_1^0 
\end{array}
\right),
\eeq
where we have suppressed the Yukawa couplings. The rows count the
singlet anti-quarks, and the columns count the doublet quarks.
The charged lepton matrices have the same form as the $Q_{em}=-1/3$
quark mass matrices.  
It is obvious that Det$M^{(2/3)}=0$, implying the massless u-quark.
But, it does not explain why the other first
family members are light. Note that the $R_U$-parity prevents the 
proton from decaying into ordinary particles such as
to $e^++\pi^0$. However, the proton can decay if
gravitino or axino is lighter than proton \cite{ccl}. In this case,
$\lambda''_{ij}$ should be severely constrained by some reasons.

\section{Conclusion}

We have explored the phenomenological constraints of the
$R_F$ parity model which gives $m_u\simeq 0$ consistent with
the strong CP solution. In addition, we put an ansatz for
the hierarchical fermion mass matrices.
Within this framework, we first required that the $R_F$ parity
model describes the charged lepton and quark masses and mixing
angles properly. In particular, from the observed down type
quark mass matrix we derived a plausible form for the sneutrino
VEV's
\beq
m_{\tilde d_{Rj}}\langle\tilde\nu_{\mu,\tau}\rangle
\gtrsim \ 400\ (\rm GeV)^2\, .
\eeq
Then we applied this constraint to
the recent neutrino oscillation data. We find that without a
sterile neutrino a large mixing angle solution of
$\nu_\mu$ and $\nu_\tau$ is not possible in the $R_F$ parity
model. However, introducing a sterile neutrino, it is possible
to generate a large mixing angle solution, if the parameter
$M_S$ is of order intermediate scale and $\langle\nu_\mu\rangle$
is of order GeV. 

Since the hierarchical fermion mass matrix in the $R_F$
parity model is given in Eq.~(\ref{mm}), we introduced a
horizontal $U(1)_X\times U(1)_Z$ symmetry such that the
fermion mass matrix arises naturally. 
This kind of horizontal symmetry can appear in string models
frequently. We also pointed out another kind of
$R$ parity, $R_U$ parity, rendering up quark massless,
but its phenomenological consequences are not explored.

\bigskip

\section*{ Acknowledgments}
J.E.K. and B.K. are supported in part by KOSEF, MOE through
BSRI 98-2468, and J.E.K. is also supported by Korea Research Foundation. 

\section*{appendix}
In this appendix, we present the interactions  which are neutral
under $U(1)_X \times U(1)_Z$ horizontal symmetry. These are
given by:

\beqar
\bullet ~~~~&& QD^cH_1\ \ {\rm type} \nonumber \\ 
&&\Theta_1^{7}\Theta_3^2~Q_3D^c_3H_1\, ,  ~~~
~\Theta_1^{12}\Theta_2^3~Q_2D^c_3H_1\, , ~~~
\Theta_1^{10}\Theta_{3}^{8}~Q_3D^c_2H_1\, , 
\nonumber \\
&&\Theta_1^{12}\Theta_3^{3}~Q_2D^c_2H_1\, ,~~~
\Theta_1^{13}\Theta_3^{8}~Q_3D^c_1H_1\, ,~~~
\Theta_1^{15}\Theta_3^{3}~Q_2D^c_1H_1\, . 
\\
\nonumber \\ \bullet ~~~~&& LE^cH_1\ \ {\rm type} \nonumber \\ 
&&\Theta_1^{7}\Theta_3^2~L_3E^c_3H_1\, ,  ~~~
~\Theta_1^{12}\Theta_2^3~L_2E^c_3H_1\, , ~~~
\Theta_1^{10}\Theta_{3}^{8}~L_3E^c_2H_1\, , 
\nonumber \\
&&\Theta_1^{12}\Theta_3^{3}~L_2E^c_2H_1\, ,~~~
\Theta_1^{13}\Theta_3^{8}~L_3E^c_1H_1\, ,~~~
\Theta_1^{15}\Theta_3^{3}~L_2E^c_1H_1\, . 
\\
\nonumber \\ \bullet ~~~~&& QU^cH_2\ \ {\rm type} \nonumber \\ 
&&Q_3U^c_3H_2\, ,  ~~~~~~~~~~~
\Theta_1^7\Theta_2^5Q_2U^c_3H_2\, ,~~~
\Theta_1^3Q_3U^c_2H_2\, , 
\nonumber \\
&&\Theta_1^{10}\Theta_2^5Q_2U^c_2H_2\, , ~~~
\Theta_1^6Q_3U^c_1H_2\, ,~~~
~~~~\Theta_1^{13}\Theta_2^5Q_2U^c_1H_2\, . 
\\ 
\nonumber \\ \bullet ~~~~&& LQD^c\ \ {\rm type} \nonumber \\ 
&&\Theta_1^{6}\Theta_3^3    L_3Q_1D^c_1\, ,  ~~~
~\Theta_1^{3}\Theta_3^3    L_3Q_1D^c_2\, ,~~~
\Theta_1^{3}\Theta_2^3    L_3Q_1D^c_3\, , 
\nonumber \\
&&\Theta_1^{10}\Theta_{2}^{2}L_2Q_1D^c_1\, , ~~~
\Theta_1^{7}\Theta_{2}^{2}L_2Q_1D^c_2\, , ~~~
\Theta_1^{10}\Theta_{2}^{8}L_2Q_1D^c_3\, ,  
\nonumber \\
&&\Theta_1^{10}\Theta_{2}^{2}L_1Q_2D^c_1\, ,  ~~~
\Theta_1^{7}\Theta_{2}^{2}L_1Q_2D^c_2\, ,  ~~~
\Theta_1^{10}\Theta_{2}^{8}L_1Q_2D^c_3 \, , 
\nonumber \\
&&\Theta_1^{6}\Theta_{3}^{3}L_1Q_3D^c_1 \, , ~~~
~\Theta_1^{3}\Theta_3^{3}L_1Q_3D^c_2 \, ,~~~
\Theta_1^{3}\Theta_2^{3}L_1Q_3D^c_3\, . 
\\
\nonumber \\ \bullet ~~~~&& LLE^c\ \ {\rm type} \nonumber \\ 
&&\Theta_1^{6}\Theta_3^3    L_1L_3E^c_1\, ,  ~~~
~\Theta_1^{3}\Theta_3^3    L_1L_3E^c_2\, ,~~~
\Theta_1^{3}\Theta_2^3    L_1L_3E^c_3\, , 
\nonumber \\
&&\Theta_1^{10}\Theta_{2}^{2}L_1L_2E^c_1\, ,  ~~~
\Theta_1^{7}\Theta_{2}^{2}L_1L_2E^c_2\, ,  ~~~
\Theta_1^{10}\Theta_{2}^{8}L_1L_2E^c_3\, .   
\eeqar
Note that the $U^cD^cD^c$ type terms are not allowed,
which ensures that the proton stability. 

Note that the so-called $\mu$ term, $\mu H_1 H_2$, 
is not allowed in the model presented above. However,
we need an electroweak scale $\mu$ term toward a successful
$SU(2)\times U(1)$ symmetry breaking~\cite{mu}. One example for the
$\mu$ term is through $\Phi_1 H_1 H_2$ where $\Phi_1$
represents the color singlet superfields introduced to cancel anomalies in
Sec. IV. For example, one can introduce the appropriate $\mu$ term by
taking $Q_{_X}(\Phi_1)=-11/3$, $Q_{_Z}(\Phi_1)=11$ and
$\vev{\Phi_1}=\mu$. This kind of charge assignment does not generate
the dangerous term $U^c D^c D^c$, but can generate the 
$L_1 H_2$ term
\beq
\Theta_2^7 \Theta_3^2 \Phi_1 L_1 H_2 = \lambda^3 \mu L_1 H_2
\approx\frac{\mu}{100} L_1 H_2,
\eeq
Note that the relative size of the induced $\mu_1$ to $\mu$
is independent of the charges of
$\Phi_1$ since the factor $\lambda^3$ is given by the differences between
charges of $H_1$ and $L_1$.
This size of the induced $\mu_1$ could break the relation $\Lambda_\tau
\gg \Lambda_e$, which we needed to avoid the present mass bound on 
$m_{\nu_e}$.  Therefore, to suppress $\Lambda_e$, we require
some cancellation between $\ve$ and  $\lambda^3 v_1$ or $\mu_1$ 
and the induced $\mu_1$.

To give VEV to $\Phi_1$ field, one can introduce 
a field $\Phi_2$ with opposite
$U(1)_X \times U(1)_Z$ charge,
$Q_{_X}(\Phi_2)=+11/3$, $Q_{_Z}(\Phi_2)=-11$ and
then, the relevant
superpotential and scalar potential are given by
\beqar
W &=& f_{\Phi}\Phi_1 H_1H_2 + M_{\Phi} \Phi_1 \Phi_2,
\nonumber \\
V &=& M_1^2 |\Phi_1|^2+M_2^2 |\Phi_2|^2
+(A_{\Phi} \Phi_1 H_1 H_2 + 
B_{\Phi} M_{\Phi} \Phi_1 \Phi_2 + {\rm H.c.} ),
\eeqar
where $M_1^2=M_{\Phi_1}^2+M_{\Phi}^2$,
$M_2^2=M_{\Phi_2}^2+M_{\Phi}^2$, and
$M_{\Phi_1}^2$, 
$M_{\Phi_2}^2$, 
$A_{\Phi}$ and $B_{\Phi}$ 
are the soft SUSY breaking parameters. 
Running $M_1^2$ can be driven negative by the
large Yukawa coupling $f_{\Phi}$. This radiative breaking
which is the same mechanism
as the radiative electroweak symmetry breaking,
generates the VEV's for $\Phi$ fields as 
$\vev{\Phi_1}=\vev{\Phi_2}\sim M_{\Phi}$ \cite{rewsb}.

\begin{figure}[ht]
\centerline{\epsfig{file=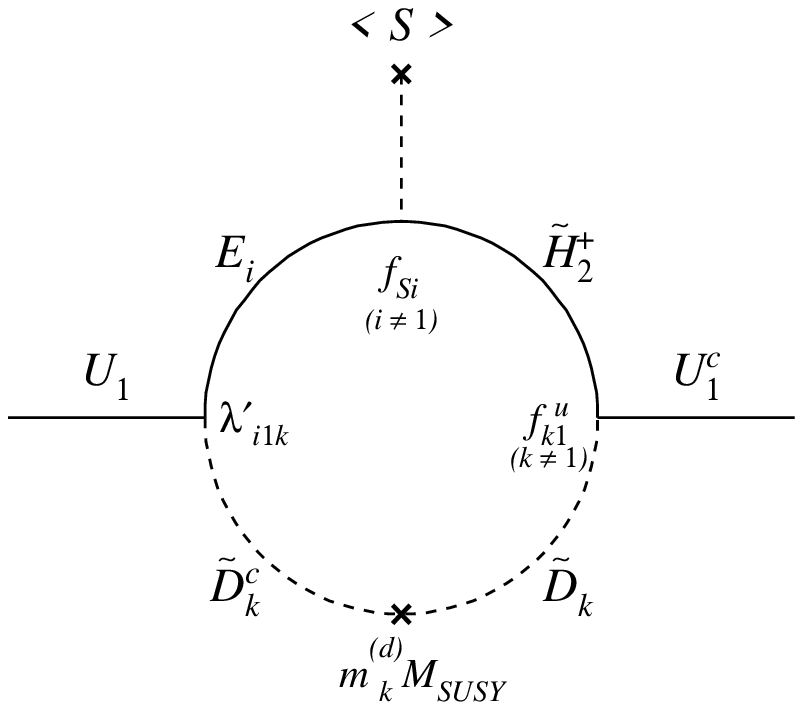,height=20cm, width=20cm}}
\caption{The one--loop $u$ quark mass.
}
\label{radum}
\end{figure}

\end{document}